\documentclass[conference]{IEEEtran}
\makeatletter
\def\ps@headings{%
\def\@oddhead{\mbox{}\scriptsize\rightmark \hfil \thepage}
 \def\@evenhead{\scriptsize\thepage \hfil \leftmark\mbox{}} 
\def\@oddfoot{} \def\@evenfoot{}} \makeatother 

\usepackage{epsfig}
\usepackage{balance}
\usepackage{subfigure}
\usepackage{rotating}
\usepackage{multirow}
\usepackage{graphicx}
\usepackage[all]{xy}

\usepackage{algorithm}
\usepackage{algpseudocode}

\newcommand{\newspacing}{\baselineskip=1.0\normalbaselineskip}

\begin{document}
%
%

\title{An Immuno-Inspired Approach to Misbehavior Detection in Ad Hoc Wireless Networks}
\date{}


\author{\IEEEauthorblockN{Martin Drozda$^\dagger$, Sebastian Schildt$^\ast$, Sven Schaust$^\dagger$ and Helena Szczerbicka$^\dagger$}
\IEEEauthorblockA{$^\dagger$ Simulation and Modeling Group, Computer Science Dept., Leibniz University of Hannover\\ 
Welfengarten 1, 30167 Hannover, Germany.}
\IEEEauthorblockA{$^\ast$ Institute of Operating Systems and Computer Networks, Technische Universit\"at Braunschweig\\
M\"uhlenpfordtstr. 23, 38106 Braunschweig, Germany.}
\IEEEauthorblockA{Email: \{drozda,svs,hsz\}@sim.uni-hannover.de, schildt@ibr.cs.tu-bs.de}}

\maketitle

\newspacing
\begin{abstract}
We propose and evaluate an immuno-inspired approach to misbehavior detection in ad hoc wireless networks. Node misbehavior can be the result of an intrusion, or a software or hardware failure. Our approach is motivated by co-stimulatory signals present in the Biological immune system. The results show that co-stimulation in ad hoc wireless networks can both substantially improve energy efficiency of detection and, at the same time, help achieve low false positives rates. The energy efficiency improvement is almost two orders of magnitude, if compared to misbehavior detection based on watchdogs. 
 
We provide a characterization of the trade-offs between detection approaches executed by a single node and by several nodes in cooperation. Additionally, we investigate several feature sets for misbehavior detection. These feature sets impose different requirements on the detection system, most notably from the energy efficiency point of view.  
\end{abstract}

\section{Introduction}

Ad hoc wireless networks can be subject to a large variety of attacks or intrusions. These attacks can range from a simple packet dropping to advanced attacks executed in collusion, possibly utilizing a superior computational platform than that of the attacked network. It is an ambition of secure protocols to prevent all or a majority of these attacks. Experience from the Internet, however, points out that flaws in these protocols are continuously being found and exploited~\cite{yegneswaran2003iig}. 

Performance analysis of security and protection solutions for ad hoc wireless networks received of a lot of interest from the community; see~\cite{mishra2004idw,zhang2003idt} for a review. It is currently unclear to what extent ad hoc wireless networks will be subject to various attacks. The history of security of home  and small mobile computing platforms however points out that such attacks can disrupt or even completely interrupt the normal operations of networks~\cite{szor2005acv}. 

In the future protecting ad hoc wireless networks can become as challenging a task as protecting home computing platforms. Many ad hoc networks are expected to be based on wireless devices with restricted computational and communication capabilities, and very limited battery resources. In many application scenarios, attack signature updates from a centralized site are infeasible. Correcting the consequences of some failures or attacks might only be possible by a costly human intervention, or not at all. An example where such a correction might be nearly impossible are underwater networks~\cite{heidemann2006rca}.

The above facts establish the basic motivation for designing autonomous detection and response systems that aim at offering an additional line of defense to the employed secure protocols. Such systems should provide several layers of functionality including the following
: (i) distributed self-learning and self-tuning with the aspiration to minimize the need for human intervention and maintenance, (ii) active response with focus on attenuation and possibly elimination of negative effects of misbehavior on the network.

The ability of self-learning and self-tuning implies a set of observable features, from which it can be deduced whether a member of an ad hoc network misbehaves or whether the conditions in the given ad hoc network could lead to decreased Quality of Service to others. An important task is therefore to identify a set of features that are useful in detecting a larger set of misbehavior types.  

Best current practices for misbehavior detection in ad hoc wireless networks are almost exclusively done on a domain knowledge basis; see~\cite{mishra2004idw,zhang2003idt,anantvalee:sid} and references therein. Although such an approach allows to find a good predictor for a specific type of misbehavior, it fails to deliver a broader knowledge on design specifics of misbehavior detection systems.    

Our goal was to benefit from the wealth of information available at the various layers of the OSI protocol stack and to provide a performance assessment of misbehavior detection done by a single node or by several nodes acting in a cooperative manner complemented with exchange of network measurements (features). We divided the examined feature set into several subsets with respect to their energy efficiency and protocol assumptions. We employed a wrapper method~\cite{kohavi1997wfs} to assess the efficiency of individual features subsets. Most importantly, we investigated applicability of a mechanism inspired by the efficiency of the Biological immune system (BIS). The BIS and its technical counterpart, Artificial immune systems (AIS), are currently under increased examination in the area of wireless networks security~\cite{sarafijanovic2004ais,kim2005dud,drozda2007mdw}. Our approach is motivated by the mechanisms that allow the key players of the BIS such as T-cells, B-cells, dendritic cells, macrophages etc. to communicate with each other and thus provide a more robust mechanism for detecting and eliminating foreign agents such as viruses or bacteria. Our focus stays especially on the remarkable ability of the BIS to avoid false positives in the classification process~\cite{murphy2007jsi}. This is demonstrated by the rareness of severe auto-immune reactions in humans (although milder forms of allergies are unfortunately rather frequent).

\smallskip
This document is organized as follows. In Section~\ref{sec:bis} we give a short overview of the Biological immune system. Section~\ref{sec:related_work} offers a summary of the related work. In Section~\ref{sec:wrapper} our evaluation approach is introduced. In Section~\ref{sec:ad_hoc} we summarize the assumptions and protocols relevant to our experiments. Section~\ref{sec:features} defines the features used in the performance evaluation. In Section~\ref{sec:arch} our immuno-inspired architecture is introduced and analyzed. Section~\ref{sec:setup} describes in detail the experimental setup. In Section~\ref{sec:results} we discuss the obtained results. And finally, in Section~\ref{sec:conclusions} we conclude and give an outlook on future research.

\section{The Biological Immune System}
\label{sec:bis}

The Biological immune system (BIS)~\cite{murphy2007jsi} can quickly recognize the presence of foreign microorganisms in the human body. It is remarkably efficient, most of the time, in correctly detecting and eliminating pathogens such as viruses, bacteria, fungi or parasites, and in choosing the correct immune response. When confronted with a pathogen, the BIS relies on the coordinated response from both of its two vital parts:

\begin{itemize}
\item the {\em innate system}: the innate immune system is able to recognize the presence of a pathogen or tissue injury, and is able to signal this to the adaptive immune system. 

\item the {\em adaptive system}: the adaptive immune system can develop during the lifetime of its host a specific set of immune responses.
\end{itemize}

For an immune reaction to occur, it is necessary that (i) a cell has been classified as a pathogen and (ii) this cell could cause some damage to the human organism. This means that the BIS is only reactive with {\em infectious} cells, i.e. with pathogens that can indeed cause harm~\cite{murphy2007jsi}. 

This demonstrates that a two-way communication, hereafter referred to as {\em co-stimulation}, between the innate and adaptive immune systems is common. Immunologists such as Frauwirth and Thompson describe co-stimulation as the involvement of  \emph{''reciprocal and sequential signals between cells''} in order to fully activate a lymphocyte~\cite{10.1172/JCI14941}. The role of lymphocytes is to recognize a specific pathogen, to trigger a corresponding immune reaction, in some forms they are also capable of pathogen elimination.

In the subsequent sections we will introduce and evaluate an approach inspired by the interplay between the innate and adaptive immune system. The goal of this approach is to help suppress false positives and at the same time achieve energy efficiency.

\section{Related Work}
\label{sec:related_work}

\subsection{AIS Based Misbehavior Detection}

The early work in adapting the BIS to networking has been done by Stephanie Forrest and her group at the University of New Mexico. In one of the first BIS inspired works, Hofmeyr and Forrest~\cite{hofmeyr1999ida} described an AIS able to detect anomalies in a wired TCP/IP network. Co-stimulation was in their setup done by a human operator who was given 24 hours to confirm a detected attack.

Sarafijanovi\'c and Le~Boudec~\cite{sarafijanovic2004ais} introduced an AIS for misbehavior detection in mobile ad hoc wireless networks. They used four different features based on the network layer of the OSI protocol stack. They were able to achieve a detection rate of about 55\%; they only considered simple packet dropping with different rates as misbehavior. A co-stimulation in the form of a danger signal emitted by a connection source was used to inform nodes on the forwarding path about perceived data packet loss. 

An AIS for sensor networks was proposed by Drozda et al. in~\cite{drozda2007mdw}. 
The implemented misbehavior was packet dropping; the detection rate was about 70\%.

Classification techniques proposed in~\cite{hofmeyr1999ida,sarafijanovic2004ais,sarafijanovic2005ais,drozda2007mdw} were based on the negative selection, a learning mechanism applied in training and priming of T-cells in the thymus. In the computational approach to negative selection due to D'haeseleer et al.~\cite{dhaeseleer1996iac}, a complement to an $n$-dimensional vector set is constructed. This is done by producing random vectors and testing them against vectors in the original vector set. If a random vector does not match anything in the original set, it becomes a member of the complement (detector) set. The vectors from the detector set are then used to identify anomalies (misbehavior). Only very recently, an efficient algorithm for negative selection was presented by Elberfeld and Textor~\cite{elbe09eas}. 


An approach based on the Danger theory
was proposed by Kim et al. in~\cite{kim2005dud}. Several types of danger signals, each having a different function are employed in order to detect routing manipulation in sensor wireless networks. The authors did not undertake any performance analysis of their approach.

It is beyond the scope of this document to provide an exhaustive summary of AIS based approaches. A review of the theoretical aspects of several BIS inspired algorithms was compiled by Timmis et al. in~\cite{timmis2008taa}. The various application areas of BIS inspired approaches were reviewed by D. Dasgupta in \cite{dasgupta2006aais}. The applicability with respect to ad hoc networks was reviewed in~\cite{drozda2010iik}.

Even though the BIS seems to be a good inspiration for improving misbehavior detection in ad hoc and sensor networks, approaches based on machine learning and similar methods received much more attention; see~\cite{anantvalee:sid,mishra2004idw} and the references therein. Despite recent efforts, {\em energy efficient} misbehavior detection remains to be an open problem. In the following sections, we use the watchdog approach due to Marti et al.~\cite{marti2000mrm} as a basis in our misbehavior detection and energy efficiency evaluation. Even though, the watchdog approach is not energy efficient, it received a great attention in the literature and it is thus perceived by many as the standard approach.


\subsection{Cascading Classifiers}

The misbehavior classification approach, that we herein present and analyze, resembles the ``cascading classifiers'' of  Kaynak and Alpaydin introduced in~\cite{kaynak2000multistage}. Their approach is based on a sequential application of several classifiers such that {\em "at the next stage, using a costlier classifier, we build a more complex rule to cover those uncovered patterns of the previous stage"}. Our BIS inspired approach can be seen as an instance of cascading classification. It is however not the complexity of classification rules that is increased at each step but the energy cost connected with observing additional states and events necessary for a more precise reasoning about a possible misbehavior.  


Cascading classifiers were empirically studied by Gama and Brazdil in~\cite{gama2000cascade}. They combined several types of classifiers: Bayes classifier, C4.5 and linear discriminant function. 
Their focal point was to investigate whether cascading classification could offer some classification performance improvement over the classification using a single classifier.




\begin{figure}[!!!t]
\begin{center}
   {
     \epsfig{file=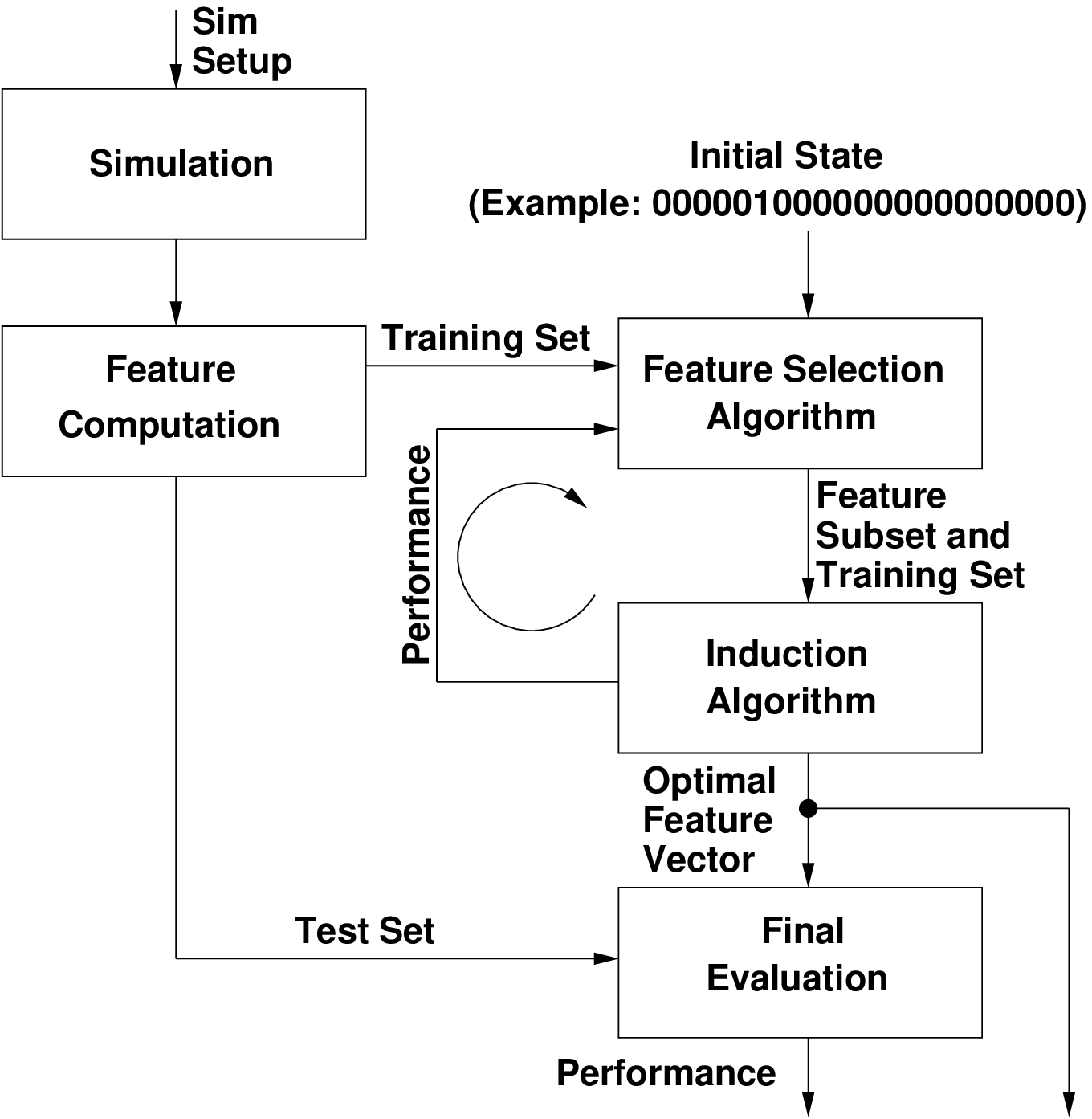, width=0.9\linewidth}
   }
 \caption{Wrapper approach.}
 \label{fig:wrapper}
 \end{center}
 \end{figure}

\section{Evaluation Approach}
\label{sec:wrapper}

To evaluate the performance of our immuno-inspired architecture, we use the wrapper approach described in~\cite{kohavi1997wfs}. The wrapper approach requires a {\em training} and a {\em test} data set as input. In our case, these data sets are obtained through simulating an ad hoc network. Our choice of network simulator was JiST/SWANS~\cite{barr2005jea}. SWANS is a Java based network simulator that offers a substantial simulation performance speed-up over other alternatives, most notably Glomosim
and ns2.
Vectors in both sets are labeled as either representing a normal behavior or behavior when the network is subject to an attack or intrusion. 

Since the feature space can be large, it is desirable to identify features that can significantly contribute to misbehavior detection. This is done by running an optimization algorithm over the feature space. The goal of the optimization is to find a (small) feature subset that is efficient in detecting misbehavior. 
In our case, the input to the optimization algorithm is encoded as a bit vector of length $k$; a bit at position $i$ set to $1$ means that the $i$-th feature is in the feature subset currently evaluated.
The wrapper approach in our setup is depicted in Fig.~\ref{fig:wrapper}. 
It can be summarized as follows:
\begin{enumerate}
\item Training and test data sets are produced through simulation.

\item An optimization algorithm computes a subset of the feature space that is expected to be optimal with respect to one or several performance measures. The optimization algorithm is initialized with an empty (with all bits set to zero) or random bit vector.

\item An induction algorithm uses a part of the training set to learn a classifier. This classifier is applied to the remaining part (hold-out set); performance measures are computed. The optimization algorithm is supplied with performance  measures of the classifier using the current feature subset. Steps (2) and (3) are repeated until a termination condition is met.  


\item At last, a thorough evaluation of the optimal feature subset using the same induction algorithm as in Step (3) and employing the test set is done. This optimal feature set alongside with its expected performance are being output. 
\end{enumerate}

The training and test sets and the hold-out set in Step (3) were produced by using stratified $n$-fold cross-validation~\cite{alpaydin2004iml}.

\section{Protocols and Definitions}
\label{sec:ad_hoc}

\subsection{Protocols}

We now state several protocols, mechanisms and assumptions relevant to our experiments. 
Node misbehavior can be the result of an intrusion,  or a software or hardware failure. Additional faults in ad hoc networks can be introduced by mobility, signal propagation, link reliability and other factors. The reason for nodes (possibly fully controlled by an attacker) to execute any form of misbehavior can range from the desire to save battery power to the intention of making an ad hoc network non-operational. 

We consider AODV~\cite{perkins99aho}, a well-known on-demand routing protocol using the RREQ and RREP handshake to establish routes, as the underlying routing protocol.

At the MAC (Medium access control) layer, the contention based IEEE 802.11 MAC protocol using both carrier sensing and RTS-CTS-DATA-ACK handshake is considered. Should the medium not be available or the handshake fails, an exponential back-off algorithm is used. This is combined with a mechanism that makes it easier for neighboring nodes to estimate transmission durations. This is done by an exchange of duration values and their subsequent storing in a data structure known as Network allocation vector (NAV). 

Alternative MAC protocols like the 802.15.4 MAC protocol
avoid using the RTS-CTS-DATA-ACK handshake, only relying on carrier sensing to access the medium, in order to decrease energy consumption at nodes.

User Datagram Protocol (UDP) is a transport layer protocol that does not guarantee any reliability, data integrity or data packet ordering.  

In {\em promiscuous mode}, a node listens to the on-going traffic among other nodes in the neighborhood and collects information from the overheard packets. Promiscuous mode is energy inefficient because it prevents the wireless interface to operate in sleep mode, forcing it into either idle or receive mode; there is also extra overhead caused by analyzing all overheard packets. According to~\cite{feeney2001iec}, power consumption in idle and receive modes is about 12-20  higher than in sleep mode. 

We do not assume any time synchronization among nodes. We assume that packets are authenticated, i.e. the sender of any packet can be easily identified and as well can be any changes in the packet body. This is a reasonable assumption in line with e.g. the ZigBee specification~\cite{zigbee}. 

\subsection{Performance Measures}

We evaluate the misbehavior classification performance in terms of detection rate and false positives (FP) rate. We assume that a classifier $\mathcal{K}$ computed by a learning algorithm is used in the classification process. The classifier $\mathcal{K}$ is then used to classify the objects $\Omega = \{o_1,..., o_p\}$, where $p$ is the number of objects. The two measures are then computed as follows:
\begin{equation}
det.~rate^{\Omega}_{c_j} (\mathcal{K}) = \frac{c_{c_j}}{n_{c_j}} \times 100.0\%
\end{equation}
\begin{equation}
FP~rate^{\Omega}_{c_j} (\mathcal{K}) = \frac{FP_{c_j}}{FP_{c_j} + c_{c_j}} \times 100.0\%
\end{equation}

where $c_j$ is the $j$-th class. $n_{c_j}$ is the number of objects labeled with the class~$c_j$; note that $n_{c_j} > 0$ in all our experiments. $c_{c_j}$ is the number of objects that were correctly classified by the learning algorithm as belonging to the class~$c_j$. $FP_{c_j}$ is the number of objects incorrectly predicted as belonging to $c_j$. 95\% confidence intervals ($CI_{95\%})$ were computed for each measure.

The overall misclassification rate is evaluated by means of the classification error:
\begin{equation}
class.~error^{\Omega} (\mathcal{K}) = \frac{\sum_{c_j} FP_{c_j}}{\sum_{c_j} n_{c_j}} \times 100.0\%
\end{equation}

\section{The Features}
\label{sec:features}

We considered 24 features from three layers of the OSI protocol stack: data link (MAC), network and transport layer. Our focus was on two basic types of features: (i) performance related such as e.g. latency or throughput and (ii) network topology related such as e.g. node degree, network diameter, average path length to a destination as recorded in the routing table. We also included features well known previously such as the watchdog feature~\cite{marti2000mrm}; some others were motivated by the results published in~\cite{zhang2003idt,drozda2007mdw}. The features, not adapted from~\cite{marti2000mrm,zhang2003idt,drozda2007mdw}, were found by considering the used protocols and choosing such features that do not add too much computational overhead. There was no formal method (Petri nets etc.) applied in this process.

Let ${s_s, s_1,...,s_i, s_{i+1},s_{i+2},..., s_d}$ be the path between $s_s$ and $s_d$ determined by a routing protocol, where $s_s$ is the source node, $s_d$ is the destination node. The features in the below introduced feature set~$f$ are averaged over a sliding time window of the size $win.~size$. Let $pcts_{TX}$ and $pcts_{RX}$ be the number of data packets sent and received by $s_i$ in a time window of the size $win.~size$, respectively.

\begin{enumerate}
\item[] {\em MAC Layer Features:}
\item[{\em M1}] {\bf MAC handshake ratio:} Computed by $s_i$ as:

\[
M1 = \frac{\sum_{p = 1}^{pcts_{TX}} \frac{\#ACK_p}{\#RTS_p}}{pcts_{TX}}
\]

where $\#RTS_p$ is the number of RTS packets sent to $s_{i+1}$ and $\#ACK_p$ is the number of ACK packets received by $s_i$ from $s_{i+1}$, when reserving the wireless medium for the packet $p$. This feature estimates the medium congestion level from the number of handshakes that were brought to completion, i.e. did not end up before an ACK was received.

\item[{\em M2}] {\bf Back-off level index:} MAC protocol back-off level $BO$ just before a data packet $p$ is transmitted from $s_i$ to $s_{i+1}$. 

\[
M2 = \frac{\sum_{p = 1}^{pcts_{TX}} BO_p} {pcts_{TX}}
\]

$M2$ is similar to $M1$, the congestion level is, however, estimated from the number of back-offs.

\item[{\em M3}] {\bf Forwarding index (watchdog):} Ratio of data packets sent from $s_i$ to $s_{i+1}$, $TX_{s_i}$  and then subsequently forwarded to $s_{i+2}$, $TX_{s_{i+1}}$.

\[
M3 = \frac{TX_{s_{i+1}}} {TX_{s_{i}}}
\]

The data packets that have $s_{i+1}$ as the destination node are excluded from this statistic.

\item[{\em M4}] {\bf Processing delay:} Time delay that a data packet accumulates at $s_{i+1}$ before being forwarded to $s_{i+2}$. 

\[
M4 = \frac{\sum_{p = 1}^{pcts_{TX}} delay_{s_{i+1}}} {pcts_{TX}}
\]

The data packets that have $s_{i+1}$ as the destination node are excluded from this statistic.

\item[{\em M5}] {\bf Data rate index:} Amount of data (in bits/s) forwarded by node $s_i$ in a time window. 

\[
M5 = \frac{\sum_{p = 1}^{pcts_{TX}} size(p)} {win.~size}
\]

where $size(p)$ is the size of a data packet $p$ that is forwarded in the given time window. The data packets that originate at $s_i$ are excluded.

\item[{\em M6}] {\bf Node degree index:} Number of neighboring nodes with which $s_i$ had an active data exchange:  
\[
M6 = \#neigh
\]  

where $\#neigh$ is the number of neighbors in a time window. 

\item[{\em M7}] {\bf 2-hop neighborhood index:} Number of neighboring nodes within 2-hop distance from $s_i$:

\[
M7 = \#2neigh
\]   

where $\#2neigh$ is the number of unique MAC layer destinations  extracted from overheard data packets in a time window. 


\smallskip
\item[] {\em Routing Layer Features:}
\item[{\em R1}] {\bf Forwarding index for RREQ:} Number of unique RREQs (i.e. with a unique source id and sequence number) forwarded by the node $s_i$. Normalized by the time window size:

\[
R1 = \frac{\#RREQ}{win.~size}
\]


\item[{\em R2}] {\bf Forwarding index for RERR:} Number of RERRs forwarded by the node $s_i$. Normalized by the time window size:

\[
R2 = \frac{\#RERR}{win.~size}
\]

\item[{\em R3}] {\bf Forwarding index for RREP:} The same ratio as in M3 but computed separately for RREP routing packets.

\item[{\em R4}]  {\bf Processing delay for RREP:} The same delay as in M4 but computed separately for RREP routing packets. 

\item[{\em R5}] {\bf Average distance to destination:} Average number of hops from $s_i$ to any destination as recorded in the routing table: 

\[
R5 = \frac{\sum_{k=1}^{table\_size} |r_k|}{table\_size}
\]

where $|r_k|$ is the length of the route $r_k$ to a destination and $table\_size$ is the number of destinations recorded.

\item[{\em R6}] {\bf Routing activity index:} Number of RREQ, RREP and RERR packets received by node $s_i$. Normalized by the time window size.

\[
R6 = \frac{\#RREQ + \#RREP + \#RERR}{win.~size}
\]

\item[{\em R7}] {\bf Connectivity index 1:} Number of unreachable destinations $\#unreach$ as recorded in the routing table of node $s_i$.

\[
R7 = \#unreach
\]

\item[{\em R8}] {\bf Connectivity index 2:} Number of invalid routes $\#invalid$ as recorded in the routing table of node $s_i$.

\[
R8 = \#invalid
\]

\item[{\em R9}] {\bf Connectivity index 3:} Number of destinations $\#dest$ with known routes as recorded in the routing table of node~$s_i$.

\[
R9 = \#dest
\]

\item[{\em R10}] {\bf Cut index 1:} Number of connections $\#connect$ with data packets forwarded by node $s_i$ in a time window.

\[
R10 = \#connect
\]

\item[{\em R11}] {\bf Cut index 2:} Number of RREP packets forwarded by node $s_i$. Normalized by the time window size.

\[
R11 = \frac{\#RREP}{win.~size}
\]

\item[{\em R12}] {\bf Diameter index:} Number of hops to the furthermost destination as recorded in the routing table of node~$s_i$. 

\[
R12 = max_k\{|r_k|\}
\]

where $k = 1 \ldots table\_size$.

\smallskip
\item[] {\em Transport Layer Features:}

\item[{\em T1}] {\bf Out-of-order packet index:} Number of DATA packets that were received by $s_i$ out of order.  

\[
T1 = \frac{\#OO}{win.~size}
\]

where $\#OO$ is the number of data packet received out-of-order. $\#OO$ is incremented, if node $s_i$ receives on the connection $c$ a data packet $p_j$ such that  $seq.~number(p_{j}) - 1 \neq seq.~number(p_{j-1})$, where $seq.~number(p_{j})$ is the sequence number of the data packet $p_j$. This assumes that the connection source uses an incremental (or similar easily predictable) scheme for computing  $seq.~number(p_{j})$. 

\item[{\em T2}] {\bf Interarrival packet delay index 1:} Average delay between data packets $p_j$ and $p_{j+1}$ sequentially received by $s_i$. The delay was computed separately {\em for each connection} and then a master average was computed.

\[
T2 = \frac{\sum_{c = 1}^{\#connect} avg\_delay_c}{\#connect}
\]

where $avg\_delay_c$ is the average delay for data packets belonging to the connection $c$. It is defined as: 
\[
\frac{ \sum_{j = 1}^{pcts^c_{RX}} delay_c(p_{j+1}, p_j)}{pcts^c_{RX}}
\] 

where $pcts^c_{RX}$ is the number of data packets received by $s_i$ on the connection $c$. $delay_c(p_{j+1}, p_j)$ is the delay between the data packets $p_{j+1}$ and $p_j$ transported by the connection $c$.

\item[{\em T3}] {\bf Interarrival packet delay variance index 1:} Variance of delay between DATA packets received by $s_i$. The variance  was computed separately for {\em each connection} and then a master average was computed.

\[
T3 = \frac{\sum_{c = 1}^{\#connect} avg\_var\_delay_c}{\#connect}
\] 

where $avg\_var\_delay_c$ is the variance of the delay for data packets belonging to the connection $c$. It is defined as:

\[
\frac{ \sum_{j = 1}^{pcts^c_{RX}} (delay_c(p_{j+1}, p_j) - avg\_delay_c)^2}{pcts^c_{RX} - 1}
\]

\item[{\em T4}] {\bf Interarrival packet delay index 2:} Average delay between DATA packets received by $s_i$. 

\[
T4 = \frac{\sum_{j = 1}^{pcts_{RX}} delay(p_{j+1}, p_j)}{pcts_{RX}}
\]

where $delay(p_{j+1}, p_j)$ is the delay between any two data subsequent packets $p_{j+1}$ and $p_j$ received by $s_i$.

\item[{\em T5}] {\bf Interarrival packet delay variance index 2:} Variance of delay between DATA packets received by $s_i$. 

\[
T5 = \frac{ \sum_{j = 1}^{pcts_{RX}} (delay_(p_{j+1}, p_j) - T4)^2}{pcts_{RX} - 1}
\]

\end{enumerate}

Occasionally, it was not possible to compute a feature because there was no traffic between $s_{i}$ and $s_{i+1}$. Such time windows were not included in our experiments. 
The features T2, T4 and T3, T5 are identical, if only data packets belonging to a single connection are received by~$s_i$.

The watchdog features M3--M4, R1--R4 and the M7 feature can be considered energy inefficient because they rely on promiscuous mode. 
Therefore, we decided to consider the following three subsets of the feature set~$f$:

\begin{enumerate}
\item $f_0 = f$.

\item $f_1 = f_0\setminus\{M3, M4, M7, R1-R4\}$. $f_1$ excludes all features that rely on promiscuous mode. 

\item $f_2 = f_1\setminus\{M1\}$. $f_2$ further excludes features based on MAC protocols using the RTS-CTS-DATA-ACK handshake. This subset could be relevant to sensor networks based on the IEEE 802.15.4 MAC protocol.
\end{enumerate}

To distinguish between a feature set and its numerical {\em instance} computed in a given time window, we introduce the following ``hat'' notation: $\hat{f}_0$, $\hat{f}_1$ and $\hat{f}_2$.

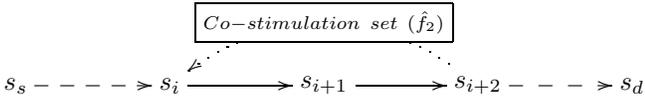
\begin{figure}[!!!t]
\begin{center}
      \xymatrix@!C@C=5pt@R=2pt{s_s \ar@{-->}[rr] && s_i \ar[rr] && s_{i+1} \ar[rr] && s_{i+2} \ar@{.>}@/_2pc/[llll]|*+[F]{{\scriptstyle Co-stimulation~set~(\hat{f}_2)}} \ar@{-->}[rr] && s_d}
\caption{Data traffic measurement model.}
\label{fig:arch}
\end{center}
\end{figure}

\section{An Immuno-Inspired  Approach}
\label{sec:arch}

\subsection{General Description}

Our approach is inspired by the co-stimulation mechanism of the BIS. Since co-stimulation benefits from the interplay between the innate and adaptive immune systems, we decided to base misbehavior detection on two different feature sets $S_1$ and $S_2$. 
 These two feature sets serve as a basis for two interconnected classifiers $\mathcal{K}(S_1)$ and $\mathcal{K}(S_2)$. 

The classification is done by sequentially applying $\mathcal{K}(S_1)$ and $\mathcal{K}(S_2)$, where $\mathcal{K}(S_1)$ mimics the classification capability of the innate immune system and $\mathcal{K}(S_2)$ mimics the classification capability of the adaptive immune system:

\begin{equation}\label{eq:costim-notation}
\xymatrix@!C@C=75pt@R=10pt{
     \mathcal{K}(S_1) \ar[r]^{co-stimulation}  & \mathcal{K}(S_2)}
\end{equation}

Let $\xi_{S_i}$, $i = {1,2}$ be a feature cost measure that reflects all feature computation costs induced by the feature set $S_i$. Our goal was to use $S_1$ and $S_2$, such that:
\begin{equation}
\label{eq:energy}
\xi_{S_1} < \xi_{S_2}
\end{equation} 
\begin{equation}
\label{eq:error}
class.~error^\Omega(\mathcal{K}(S_1)) > class.~error^\Omega(\mathcal{K}(S_2))
\end{equation} 

This means, first a classifier with a lower feature cost but a higher classification error is applied. If the classifier $\mathcal{K}(S_1)$ detects a misbehavior, then the classifier $\mathcal{K}(S_2)$ with a higher feature cost and a lower classification error is applied. Hereafter we refer to the conditional process that triggers the $\mathcal{K}(S_2)$ classification as {\em co-stimulation}. Note that if the condition expressed in Eq.~\ref{eq:error} is not fulfilled, only a classification based on $S_1$ is necessary, since $\mathcal{K}(S_1)$ would offer both a lower classification error and a lower feature cost. 

Co-stimulation in the BIS can take form of a feedback loop~\cite{10.1172/JCI14941}, where one part of the BIS interacts with another part and vice-versa, until the activation threshold for an immune reaction is met. For the ease of co-stimulation performance analysis, we decided not to consider such a feedback loop this time.

\begin{algorithm*}[!t]
\caption{Co-stimulation based misbehavior detection at node $s_i$}\label{fig:detmodel} 
\begin{algorithmic}[1] 
\Require Sufficient data traffic in current time window
\Require $\hat{f}_2^{s_{i+2}} $ from 2-hop downstream node
\Procedure{DETECT\_MISBEHAVIOR}{}
                \State boolean $suspicious \gets False$
	   	\State $ \hat{f}_{2}^{s_i}  \gets $ \Call{compute\_$f_2$\_feature\_sample}{$s_i $}\label{alg:line0}
		
	   	\State $ \hat{\mathcal{F}}_2 \gets \hat{f}_2^{s_i } \cup \hat{f}_2^{s_{i+2}} $\medskip
		\If{ \Call{Classification}{$\hat{\mathcal{F}}_2$} $ == misbehavior $  } \label{alg:line1}
	   		\State $ suspicious \gets $ True
	   	\EndIf \medskip
	   	\If{ $suspicious $ == True }  \Comment { Co-stimulation: $\mathcal{K}(\mathcal{F}_2) \rightarrow \mathcal{K}(f_0)$ }
     
	   		\State $ \hat{f}_0 \gets $ \Call{compute\_$f_0$\_feature\_sample}{$s_i$}
			\If{ \Call{Classification}{$\hat{f}_0$} $ == misbehavior $  }  \label{alg:line2}
	   	    	\State \Call{mark\_as\_misbehaving}{$s_{i+1}$}  \Comment{Misbehavior confirmed}
	   	          \EndIf
	   	\EndIf
\EndProcedure 
\end{algorithmic} 
\end{algorithm*} 

\subsection{Detailed Description}

Co-stimulation can be in ad hoc wireless networks implemented in several ways depending on the application scenario, the applied protocols or the expected misbehavior type. Let us focus on an application scenario with the assumption that a misbehaving node can be detected by its neighbors.

This translates to our architecture as follows: the node $s_{i+2}$ computes the restricted feature set $\hat{f}_2$ (or $\hat{f}_1$). This (co-stimulatory) feature set is then proliferated in the upstream connection direction (towards connection source); see Fig.~\ref{fig:arch}. 
Since the computation of $\hat{f}_2$ is time window based, the frequency with which $\hat{f}_2$ gets sent depends on the time window size. In our implementation, $\hat{f}_2$ was sent out immediately at the end of each time window. Upon receiving $\hat{f}_2$, the node $s_{i}$ compares it with its own $\hat{f}_2$ sample (later we introduce a machine learning approach that implements this comparison task). Based on this, a behavior classification with respect to the node $s_{i+1}$ is done.  If $s_{i}$ classifies $s_{i+1}$ as misbehaving, it will compute $\hat{f}_0$. If misbehavior is detected again, $s_{i+1}$ is finally classified as misbehaving. 

Before proceeding any further, we introduce the following notation in order to keep track of composite feature sets computed at $s_{i}$ and $s_{i+2}$: $\mathcal{F}_0^{s_i} = f_0^{s_i}  \cup f_0^{s_{i+2}}$, $\mathcal{F}_1^{s_i} = f_1^{s_i}  \cup f_1^{s_{i+2}}$ and $\mathcal{F}_2^{s_i} = f_2^{s_i}  \cup f_2^{s_{i+2}}$. Similarly, $\hat{\mathcal{F}}_0^{s_i} = \hat{f}_0^{s_i}  \circ \hat{f}_0^{s_{i+2}}$, $\hat{\mathcal{F}}_1^{s_i} = \hat{f}_1^{s_i}  \circ \hat{f}_1^{s_{i+2}}$ and $\hat{\mathcal{F}}_2^{s_i} = \hat{f}_2^{s_i}  \circ \hat{f}_2^{s_{i+2}}$, where $\circ$ is the operator of vector concatenation. For simplicity, whenever clear from the context, we will omit the superscripts. 


A formal description of our co-stimulation inspired approach is presented in Alg.~\ref{fig:detmodel}. With respect to the notation introduced in Eq.~\ref{eq:costim-notation}, our co-stimulation inspired approach can be succinctly expressed as:

\begin{equation}
\xymatrix@!C@C=25pt@R=10pt{
       \mathcal{K}(\mathcal{F}_2) \ar[rr]^{co-stimulation}  && \mathcal{K}(f_0)}
\end{equation}

In the following sections, we show, using experiments and an energy model for an IEEE 802.11 wireless card, that Eq.~\ref{eq:energy} and Eq.~\ref{eq:error} hold in this case.

A few observations can be made at this stage. The proliferation of $\hat{f}_2$ can be implemented without adding any extra communication complexity by attaching this information to CTS or ACK MAC packets (a modification the MAC protocol would be necessary). As long as there are DATA packets being forwarded on this connection, the feature set can be proliferated. If there is no DATA traffic on the connection (and thus no CTS/ACK packets exchanged), the relative necessity to detect the possibly misbehaving node $s_{i+1}$ decreases. Optionally, proliferation of $\hat{f}_2$ can be implemented by increasing the radio radius at $s_{i+2}$, by broadcasting it with a low time-to-live value or by using a standalone packet type.

If the node $s_{i+1}$ decides not to cooperate in forwarding the feature set information, the node $s_{i}$ will switch, after a time-out, to $\hat{f}_0$ computation. In this respect, not receiving $\hat{f}_2$ can be understood as a form of {\em negative co-stimulation}. If the goal is to detect a node controlled by an attacker, it is important that the originator of  $\hat{f}_2$ can be unambiguously identified, i.e. an authentication is necessary. An additional requirement is the use of sequence numbers for $\hat{f}_2$. Otherwise, the misbehaving node $s_{i+1}$ could interfere with the mechanism by forwarding outdated cached $\hat{f}_2$.

In general, if $\hat{f}_0$ were proliferated instead of $\hat{f}_2$, the size of the feature space at the receiving node $s_{i}$ doubles, in our case to 48 features. If higher statistical moments were computed for all the features in $f_0$, the feature space size could easily reach (or even surpass) 100 depending on the complexity of employed protocols. This fact translates it into a typical machine learning classification problem that we discuss in the following sections.

To summarize, after $s_{i}$ receives $\hat{f}_2$: (i) it will process $\hat{f}_2$, if it is not originating from $s_{i+1}$ or (ii) it will forward $\hat{f}_2$ to all upstream neighbors, otherwise. This means, each such feature set travels only two hops. Notice that, in general, $s_{i+1}$ or $s_{i+2}$ can have several successor or predecessor nodes.  This happens if $s_{i+1}$ or $s_{i+2}$ forwards data packets for several connections. 
More formally, we assume that $|\bullet s_{i+1}| \geq 1$, $|\bullet s_{i+2}| \geq 1$ and $|s_{i+1} \bullet| \geq 1$, where $\bullet s_{k}$, $s_{k} \bullet$ are the sets of all predecessor and successor nodes of $s_{k}$, respectively. The upper limit for the number of received $\hat{f}_2$ per time window is thus the number of connections for which $s_{i}$ forwarded data packets. 


\subsection{Co-stimulation Analysis}

Let us now take a closer look at the classification performance of our co-stimulation approach. Let $\Omega$ be a set of vectors that should be classified. Let $\Omega_{\mathcal{F}_2} \subseteq \Omega$ be the subset of vectors that were marked as suspicious (i.e. possibly representing a misbehavior), after $\mathcal{K}(\mathcal{F}_2)$ was applied. 

Let us first assume that $class.~error^{\Omega}(\mathcal{K}(f_0)) = 0$. If $\mathcal{K}(f_0)$ is applied to $\Omega_{\mathcal{F}_2}$, it clearly holds:
\begin{equation}
class.~error^{\Omega_{\mathcal{F}_2}}(\mathcal{K}(f_0)) = 0
\end{equation}

This implies, for the final FP rate with respect to a misbehavior class $c_j$, after co-stimulation is applied, it holds:
\begin{equation}
\label{eq:fp_rate}
FP~rate^{\Omega}_{c_j}(\mathcal{K}(\mathcal{F}_2) \rightarrow \mathcal{K}(f_0)) = 0
\end{equation}

In other words, the application of $\mathcal{K}(f_0)$ removes all vectors that were misclassified by $\mathcal{K}(\mathcal{F}_2)$. Furthermore, the final detection rate for the given class $c_j$ is determined by the detection rate of the $\mathcal{K}(\mathcal{F}_2)$ classifier:
\begin{equation}
\label{eq:det_rate}
det.~rate^{\Omega}_{c_j}(\mathcal{K}(\mathcal{F}_2) \rightarrow \mathcal{K}(f_0)) = det.~rate^{\Omega}_{c_j}(\mathcal{K}(\mathcal{F}_2)
\end{equation}

The rationale is, only the vectors correctly classified after the $\mathcal{K}(\mathcal{F}_2)$ classification form a basis for $\mathcal{K}(f_0)$ classification. 


Since to achieve $class.~error^{\Omega}(\mathcal{K}(f_0)) = 0$ may not be possible, Eq.~\ref{eq:det_rate} translates for $class.~error^{\Omega}(\mathcal{K}(f_0)) \neq 0$ to:
\begin{equation}
\label{eq:det_rate_final}
det.~rate^{\Omega}_{c_j}(\mathcal{K}(\mathcal{F}_2) \rightarrow \mathcal{K}(f_0)) \leq det.~rate^{\Omega}_{c_j}(\mathcal{K}(\mathcal{F}_2))
\end{equation}

To formulate a similar relationship for the false positives rate is non-trivial, since, in general, $\Omega_{\mathcal{F}_2}$ can be an arbitrary subset of $\Omega$. However, if:
\begin{equation}
FP~rate^{\Omega_{\mathcal{F}_2}}_{c_j}(\mathcal{K}(f_0)) = FP~rate_{c_j}^{\Omega}(\mathcal{K}(f_0))
\end{equation}

then it holds:
\begin{equation}
\label{eq:fp_rate_final}
FP~rate^{\Omega}_{c_j}(\mathcal{K}(\mathcal{F}_2) \rightarrow \mathcal{K}(f_0)) = FP~rate^{\Omega}_{c_j}(\mathcal{K}(f_0))
\end{equation}


The validity of this relationship can be well estimated through experimental analysis. Our experimental setup and the related results are presented in Sec.~\ref{sec:setup} and Sec.~\ref{sec:results}.

\section{Experimental Setup}
\label{sec:setup}

\smallskip
{\em Topology, Connections, Data Traffic and Protocols:} We used a topology based on a snapshot from the movement prescribed by the Random waypoint movement model~\cite{johnson1996dsr}. There were 1,718 nodes simulated; the physical area size was 3,000m $\times$ 3,000m. We used this topology due to the fact that this topology was quite extensively studied by Barrett et al. in~\cite{barrett2005upp}; results reported therein include many graph-theoretical measures that were helpful in finding suitable parameters for our experiments. 

 We modeled data traffic as Constant bit rate (CBR), i.e. there was a constant delay when injecting data packets. This constant delay in our experiments was 2 seconds (injection rate of 0.5 packet/s); the packet size was 68 bytes. CBR data packet sources correspond e.g. to sensors that transmit their measurements in predefined constant intervals. CBR can be considered an extreme model for data packet injection due to its synchronized nature (if a data packet collision at a node occurs, there is a high chance that it occurs again in the future). In fact, the results published in~\cite{schaust2008inp} show that when using a stochastic injection model, such as the Poisson traffic model, one can expect a better performance of the detection system. 

We used 50 concurrent connections. The connection length was 7 hops. In order to represent a dynamically changing system, we allowed connections to expire. An expired connection was replaced by another connection starting at a random source node that was not used previously. Each connection was scheduled to exist approximately $15$ to $20$ minutes. The exact connection duration was computed as
\begin{equation}
\delta + r_U \lambda
\end{equation}

where $\delta$ the desired duration time of a connection, $r_U$ a random number from the uniform distribution $[0,1]$ and $\lambda$ the desired variance of the connection duration. In our experiments, we used $\lambda = 5~min$.

We used the AODV routing protocol, IEEE 802.11b MAC protocol, UDP transport protocol and IPv4. The channel frequency was set to 2.4 GHz. The transmission rate was set to 2 Mbps. We used the Two-ray signal propagation model~\cite{rappaport2001wcp}. Antenna and signal propagation properties were set so that the resulting radio radius equaled 100 meters.


\smallskip
{\em Misbehavior models:} We considered three types of misbehavior. (i) DATA packet dropping: 30\% DATA packets were randomly and uniformly dropped at misbehaving nodes. (ii) DATA packet delaying: 30\% DATA packets were randomly and uniformly delayed by 0.1 second at misbehaving nodes. (iii) Wormholes~\cite{hu2006waw}. Wormholes are private (out-of-band) links between one or several pairs of nodes. They are added by an attacker in order to attract data traffic into them to gain control over packet routing and other network operations. There were 20 wormholes in each simulation run; the length of wormholes was 15 hops, i.e. the source and sink were 15 hops away before a given wormhole was activated. 

There were 236 randomly chosen nodes to execute DATA dropping or delaying misbehavior. As it is hard to predict the routing of packets, many of these nodes could not execute any misbehavior as there were no DATA packets to be forwarded by them. In our case, 236 misbehaving nodes resulted in about 20-30 actively (concurrently) misbehaving nodes. 

In case of the dropping and delaying misbehavior, our intention was to model random failure occurrences, assuming a uniform failure distribution in the network. The wormhole misbehavior is an instance of misbehavior done in {\em collusion}, i.e. two nodes must closely cooperate. Any of these types of misbehavior can have a significant impact on the medium contention resolution. DATA dropping removes packets from the network and can thus decrease medium congestion. DATA delaying impacts the distribution of medium contention. Wormholes can cause severe changes to a network's topology and therefore impact the quality of medium contention at certain nodes (such as those lying on a network cut before a wormhole was activated).   

\smallskip
{\em Experiments:} We did 20 independent runs for each misbehavior type and 20 misbehavior free (normal) runs ($4 \times 20$ runs in total). The simulation time for each run was 4 hours. We used a non-overlapping time window approach for the feature computation. We used four different time window sizes: 50, 100, 250 and 500 seconds. In case of a 500-second time window, there were 28 non-overlapping windows in each run (4 hours/500 seconds = 28.8). This gave us 4 $\times$ 20 $\times$ 28 = 2,240 vectors (samples) for each node.

\smallskip
{\em Labeling and constructing the training and test sets:} 
The vectors in the training and test sets were labeled as follows: if the node $s_{i+1}$ was in a given time window misbehaving, i.e. dropping/delaying packets or the start point of a wormhole, the vector $\hat{\mathcal{F}}_0, \hat{\mathcal{F}}_1$ or $\hat{\mathcal{F}}_2$ was labeled with the respective misbehavior class. The vectors from normal runs were all labeled as ``normal''.
In order to simplify the experiments, we only considered 20 distinct nodes with high traffic rates and different node degrees. The nodes were chosen to ensure that enough vectors representing each misbehavior class were available. Notice that some vectors were excluded from the experiments because there was no data traffic between $s_{i}$ and~$s_{i+1}$.

\smallskip
{\em Optimization algorithm:} We used \emph{forward selection} as optimization algorithm for the wrapper approach~\cite{alpaydin2004iml}. This algorithm starts with an empty feature set. The feature that decreases the residual classification error the most will then be added. Following this greedy approach, a new feature gets added as long as it decreases the residual classification error. Optimization by means of forward feature selection delivers a normalized feature weight vector. These normalized weights (range between $0.0-1.0$) are proportional to the significance of a given feature. A feature weight vector was computed for each of the top 20 nodes. An average feature weight vector is reported.


\smallskip
{\em Induction algorithm:} We used a decision tree classifier. 
We were interested in a less complex algorithm since the wrapper method requires that the algorithm is executed multiple times. 
To decide whether a node within the decision tree should be further split (impurity measure), we used the information gain measure. 
As the decision tree classifier is a well-known algorithm, we omit its discussion. We refer the interested reader to~\cite{alpaydin2004iml}.  

We used implementations of the optimization and induction algorithms from the Rapidminer tool~\cite{mierswa2006}. Rapidminer is an open-source tool for complex data mining tasks.  

Parameters in the experimental setup are summarized in Fig.~\ref{fig:parameters}.

\begin{figure*}[!!!tp]
\begin{center}
\noindent \fbox{
\begin{minipage}{17.65cm}
\small{      
\begin{enumerate}
\item {\bf Induction algorithm:} Decision tree with information gain measure (impurity measure).
\item {\bf Feature selection algorithm:} Forward feature selection.
\item {\bf Validation approach:} $n$-fold cross-validation with $n = 20$.
\item {\bf Misbehavior types:} Packet dropping, packet delaying, wormholes. 
\item {\bf Performance measures:} Classification error, detection rate and false positives rate, their arithmetic average and 95\% confidence intervals. 
\item \textbf{Network topology:} Snapshot of movement modeled by random waypoint mobility model i.e. it is a static network. There were 1,718 nodes. The area was a square of 2,900m$\times$2,950m. The transmission range of transceivers was 100 meters.
\item  \textbf{Number of connections:} 50 CBR (constant bit rate) connections.
\textbf{MAC protocol}: IEEE 802.11b DCF. \textbf{Routing protocol}: AODV. 
Other parameters: (i) Propagation path-loss
  model: two ray (ii) Channel frequency:
  2.4 GHz (iii) Topography: Line-of-sight (iv) Radio type: Accnoise (v)
  Network protocol: IPv4 (vi) Transport protocol: UDP. 
\item \textbf{Injection rate:} 0.5 packet/second. Data packet size was 68 bytes.
\item The number of independent simulation runs for each combination of input parameters was 20. The simulation time was 4 hours. 
\item \textbf{Simulator used:} JiST/SWANS; hardware used: 30$\times$ Linux (SuSE 10.0) PC with 2GB RAM and Pentium 4 3GHz microprocessor.

\end{enumerate}
}
\end{minipage}
}
\end{center}
\caption{Parameters used in the experiment.}
\label{fig:parameters}
\end{figure*}

\begin{table*}[!!!t]
\begin{center}

\begin{tabular}{|c|c|c|c|c|c|c|c|c|c|c|}
\hline
\multicolumn {1}{|c|}{} & \multicolumn {2}{|c|}{Normal} & \multicolumn {2}{|c|}{Dropping} & \multicolumn {2}{|c|}{Delaying} & \multicolumn {2}{|c|}{Wormhole} & \multicolumn {2}{|c|}{Any misbehavior}\\
\hline
Window size[s] &Det. rate& $CI_{95\%}$ &Det. rate &$CI_{95\%}$ &Det. rate &$CI_{95\%}$ &Det. rate &$CI_{95\%}$ &Det. rate &$CI_{95\%}$\\
\hline
\multicolumn {1}{|c|}{} & \multicolumn {10}{|c|}{$\mathcal{K}(\mathcal{F}_0)$}\\

\hline
50 &99.45 &0.15 &99.37 &0.17 &99.39 &0.16 &87.59 &5.84 &98.15 &0.62 \\
100 &99.45 &0.19 &99.64 &0.22 &99.51 &0.22 &84.15 &6.68 &97.75 &0.71 \\
250 &99.20 &0.28 &99.27 &0.40 &99.47 &0.26 &84.49 &4.38 &97.50 &0.66 \\
500 &98.93 &0.45 &98.81 &0.67 &99.01 &0.44 &83.63 &5.03 &97.19 &0.85 \\

\hline
\multicolumn {1}{|c|}{} & \multicolumn {10}{|c|}{$\mathcal{K}(\mathcal{F}_1)$}\\

\hline
50 &97.89 &0.39 &94.31 &1.09 &92.86 &0.89 &82.15 &5.87 &93.55 &0.97 \\
100 &97.06 &0.39 &91.68 &1.37 &88.59 &1.20 &78.74 &6.36 &90.48 &1.23 \\
250 &94.70 &0.69 &88.42 &2.55 &78.91 &1.93 &74.18 &8.23 &84.34 &1.65 \\
500 &91.59 &1.48 &84.20 &4.11 &64.00 &3.34 &71.37 &8.87 &76.03 &2.26 \\

\hline
\multicolumn {1}{|c|}{} & \multicolumn {10}{|c|}{$\mathcal{K}(\mathcal{F}_2)$}\\

\hline
50 &97.94 &0.37 &94.28 &1.10 &92.89 &0.89 &82.21 &5.91 &93.58 &0.99 \\
100 &97.33 &0.35 &91.71 &1.36 &88.53 &1.36 &78.21 &6.85 &90.30 &1.28 \\
250 &94.82 &0.66 &88.37 &2.55 &78.65 &1.87 &73.91 &8.48 &84.19 &1.49 \\
500 &91.73 &1.73 &84.13 &4.15 &64.64 &3.41 &71.16 &9.01 &76.40 &2.53 \\

\hline
\end{tabular}
\end{center}
\caption{Detection rate.}
\label{tab:perform}
\end{table*}

\begin{table*}[!!!t]
\begin{center}

\begin{tabular}{|c|c|c|c|c|c|c|c|c|c|c|}
\hline
\multicolumn {1}{|c|}{} & \multicolumn {2}{|c|}{Normal} & \multicolumn {2}{|c|}{Dropping} & \multicolumn {2}{|c|}{Delaying} & \multicolumn {2}{|c|}{Wormhole} & \multicolumn {2}{|c|}{Any misbehavior}\\
\hline
Window size[s] &~FP rate~~& $CI_{95\%}$ &~FP rate~~&$CI_{95\%}$ &~FP rate~~&$CI_{95\%}$ &~FP rate~~&$CI_{95\%}$ &~FP rate~~&$CI_{95\%}$\\

\hline
\multicolumn {1}{|c|}{} & \multicolumn {10}{|c|}{$\mathcal{K}(\mathcal{F}_0)$}\\

\hline
50 &1.04 &0.33 &0.39 &0.13 &0.78 &0.17 &4.35 &1.67 &0.94 &0.23 \\
100 &1.30 &0.41 &0.11 &0.08 &0.61 &0.26 &5.04 &1.81 &0.88 &0.28 \\
250 &1.48 &0.39 &0.34 &0.30 &0.49 &0.33 &8.32 &3.04 &1.44 &0.62 \\
500 &1.72 &0.57 &0.74 &0.77 &0.62 &0.44 &9.81 &3.75 &1.77 &0.66 \\

\hline
\multicolumn {1}{|c|}{} & \multicolumn {10}{|c|}{$\mathcal{K}(\mathcal{F}_1)$}\\

\hline
50 &3.73 &0.57 &3.46 &0.73 &5.47 &0.77 &8.12 &2.61 &3.63 &0.54 \\
100 &5.59 &0.84 &4.52 &0.98 &8.60 &0.86 &9.55 &3.35 &5.10 &0.53 \\
250 &9.02 &0.97 &6.46 &1.99 &15.61 &2.18 &16.86 &5.91 &9.41 &1.19 \\
500 &13.74 &1.63 &10.57 &3.36 &25.96 &3.05 &19.85 &6.95 &15.29 &1.76 \\

\hline

\multicolumn {1}{|c|}{} & \multicolumn {10}{|c|}{$\mathcal{K}(\mathcal{F}_2)$}\\
\hline
50 &3.71 &0.57 &3.44 &0.71 &5.28 &0.76 &8.23 &2.60 &3.55 &0.52 \\
100 &5.68 &0.86 &4.40 &0.99 &8.38 &0.91 &6.69 &2.12 &4.65 &0.44 \\
250 &9.11 &0.94 &6.22 &1.85 &15.38 &2.06 &16.05 &5.91 &9.28 &1.25 \\
500 &13.54 &1.81 &11.08 &3.73 &25.24 &3.56 &19.81 &7.52 &15.09 &2.27 \\

\hline

\end{tabular}
\end{center}
\caption{False positives rate.}
\label{tab:perform2}
\end{table*}

\section{Performance Evaluation}
\label{sec:results}

\subsection{Basic Performance}

In this subsection, we discuss the misbehavior detection performance of $\mathcal{K}(\mathcal{F}_0)$, $\mathcal{K}(\mathcal{F}_1)$ and $\mathcal{K}(\mathcal{F}_2)$. We first discuss the performance of these three classifiers in isolation. The performance of our co-stimulation based approach is discussed in Sec.~\ref{sec:costim} and Sec.~\ref{sec:costim-energy}.

The results with respect to the detection and false positives rate are reported in Table~\ref{tab:perform} and~\ref{tab:perform2}. The detection and FP rates were computed also for all types of misbehavior merged into a single class. This allows for a better understanding of ``misbehavior/no-misbehavior'' classification performance. It can be observed that $\mathcal{K}(\mathcal{F}_0)$ performed very well with respect to normal behavior, and dropping and delaying misbehavior. In case of the wormholes misbehavior, the FP rate was in the range $4.35-9.81\%$. 
Both $\mathcal{K}(\mathcal{F}_1)$ and $\mathcal{K}(\mathcal{F}_2)$ perform much worse than $\mathcal{K}(\mathcal{F}_0)$. The FP rate for $\mathcal{K}(\mathcal{F}_1)$ and $\mathcal{K}(\mathcal{F}_2)$ was in many cases alarmingly high. Decreasing the time window size delivered in many cases a statistically significant improvement.

In Table~\ref{tab:conf_matrix}, the confusion matrices for $\mathcal{K}(\mathcal{F}_0)$ and $\mathcal{K}(\mathcal{F}_2)$ with window size equal 50 seconds are shown. A confusion matrix contains information about actual and predicted classifications done by a classification algorithm. Only the classification outcome for the top 20 nodes is reported. 
Notice that sample sizes per node were not identical. Therefore, the detection and FP rates reported in Table~\ref{tab:perform} and~\ref{tab:perform2} are slightly different from the detection and FP rates that can be derived from Table~\ref{tab:conf_matrix}. It can be seen that the wormhole misbehavior was often misclassified as normal behavior. The reason for that is, according to our wormhole model, a wormhole is used only if it lies on the shortest path to a destination. This means, in some cases only a small fraction of the data traffic received at $s_{i+1}$ was taking ``advantage'' of the wormhole. Such a traffic pattern was harder to distinguish from the normal traffic. Note also that the sample size for the wormhole misbehavior is smaller than by the other classes. This is due to the fact that the number of wormholes in a network must not exceed a certain limit, otherwise the induced topological changes become extreme.

An interesting question is which features contributed the most to the overall performance. Feature weights are reported in Table~\ref{tab:perform3} (only features with a weight greater or equal $0.25$ are shown). Features that were computed locally (at node $s_i$) are appended with ``L''; features that were received from a remote node ($s_{i+2}$) are appended with ``R'' (see Fig.~\ref{fig:arch}).

$\mathcal{K}(\mathcal{F}_0)$ was dominated by the watchdog features with M3$\_$L and M4$\_$L having the highest weights. $\mathcal{K}(\mathcal{F}_2)$ with 50s time window was dominated by R9$\_$L, T1$\_$L, T3$\_$L, T1$\_$R and T3$\_$R. Packet dropping and delaying classification was based on the features T1$\_$L, T3$\_$L, T1$\_$R and T3$\_$R (we inspected the actual rules computed by the decision tree classifier). In this case, the classification was based on learning the differences in traffic at nodes $s_i$ and $s_{i+2}$. 

In the absence of the watchdog features, precision of wormhole classification was benefiting from the two topology features R5$\_$L and R9$\_$L; the reason is that wormholes increase the number of nodes that lie in a node's neighborhood. This precludes the necessity to rely on the features received from $s_{i+2}$ since under wormhole misbehavior $s_{i+2}$ might be  identical with a wormhole's end point. We would like to point out that even though with R5$\_$L can be easily manipulated by the wormhole, R9$\_$L is much more resilient. This feature reflects the increased routing utility of nodes that lie in the neighborhood of a wormhole. It should also be noted that when using $\mathcal{F}_0$, wormhole detection is partly based on the M3$\_$L watchdog feature, since wormholes by sending over a private link appear not to be forwarding packets. For the same reason, in more dynamic scenarios, the existence of a wormhole might also get exposed, if medium congestion at its start point drops significantly.  


Another result to be seen in Table~\ref{tab:perform3} is the insignificance of the MAC features M1 and M2. Both M1 and M2 indirectly measure the medium congestion around a node. This means, should a node be dropping packets and thus decreasing the need for medium access, this should get detected with the help of M1. On the other hand, any packet dropping should also get directly detected by the watchdog feature M3. There was a consideration that as M1 increases in value, M3 would accordingly decrease in value, i.e. M1 and M3 would form a kind of dynamic equilibrium. This however could not be demonstrated in our setup. The reason for that seems to be the relatively low data packet injection rate (0.5 packet/s) connected with additional data packets dropping (further decreasing the number of data packets being forwarded). We believe, studying such equilibria could improve performance of misbehavior detection, at least in more dynamical scenarios than the one presented herein.

\begin{table}[!!!t]
\begin{center}
\subfigure[$\mathcal{K}(\mathcal{F}_0)$. Avg. sample size per node = 4,096.]{
\begin{tabular}{|c|c|c|c|c|c|}
\hline
&&\multicolumn {4}{|c|}{Actual}\\
\hline
&&1&2&3&4\\
\hline
\multirow{1}{*}{\begin{sideways}\parbox{12mm}{Predicted}\end{sideways}}
&1 &51629  &66     &104    &433\\    
&2 &47     &10927  &6      &1\\
&3 &88     &5      &15847  &9\\
&4 &120    &2      &2      &2990\\
\hline
\end{tabular}}

\subfigure[$\mathcal{K}(\mathcal{F}_2)$. Avg. sample size per node = 3,789.]{
\begin{tabular}{|c|c|c|c|c|c|}
\hline
&&\multicolumn {4}{|c|}{Actual}\\
\hline
&&1&2&3&4\\
\hline
\multirow{1}{*}{\begin{sideways}\parbox{12mm}{Predicted}\end{sideways}}
&1 &50794  &565    &1177   &575\\
&2 &231    &10285  &138    &8\\
&3 &658    &139    &14620  &41\\
&4 &201    &11     &24     &2809\\
\hline
\end{tabular}}

\end{center}
\caption{Confusion matrices for (a)~$\mathcal{K}(\mathcal{F}_0)$ and (b)~$\mathcal{K}(\mathcal{F}_2)$. Window size = 50 seconds. 1=Normal, 2=Dropping, 3=Delaying, 4=Wormhole.}
\label{tab:conf_matrix}
\end{table}

\begin{table}[!!!t]
\begin{center}

\begin{tabular}{|c|c|c|c|c|}
\hline
\multicolumn {1}{|c|}{} & \multicolumn {2}{|c|}{500s} & \multicolumn {2}{|c|}{50s}\\
\hline
Feature Id & $\mathcal{F}_0$ & $\mathcal{F}_2$ & $\mathcal{F}_0$ & $\mathcal{F}_2$\\
\hline
M3$\_$L       &0.80   &&0.95   &\\
M4$\_$L       &0.95   & &1.00   &\\
M5$\_$L       &   &&           &0.25 \\
M6$\_$L       &          &0.45   &&\\
R5$\_$L&     &  &0.30           &0.55\\
R9$\_$L      &0.60           &0.40   &0.85           &1.00\\
R11$\_$L&     &0.25  &           &0.35\\
R12$\_$L && &          &0.40 \\
T1$\_$L      &          &0.65   &           &0.60 \\
T2$\_$L&     &   &     &0.25\\
T3$\_$L      &           &0.45   &0.25           &0.70\\
T4$\_$L&     &0.25   &           &0.25\\
T5$\_$L&     &0.30   &0.35           &0.50\\
M6$\_$R&&&         &0.30\\
R9$\_$R&     &   &0.30           &0.35 \\
R11$\_$R&     &   &     &0.40\\
R12$\_$R&     &   &0.25           &0.25\\
T1$\_$R      &           &0.80   &           &0.70\\
T2$\_$R&     &   &     &0.30\\
T3$\_$R&     &   &0.30           &0.95\\
T4$\_$R&     &  &     &0.35\\
T5$\_$R&     &0.65   &     &0.50\\
\hline

\end{tabular}
\end{center}
\caption{Feature weights. Window size=\{500s, 50s\}. A~blank field means that either the feature does not belong to the feature set or its value was $< 0.25$.}
\label{tab:perform3}
\end{table}

\subsection{Implications for Design of Autonomous Detection Systems}

The results in Tables~\ref{tab:perform}, \ref{tab:perform2} and \ref{tab:perform3} demonstrate the relative strength of the used watchdog features. It is clear that they can be used almost alone and their classification ability is better than the combined classification ability of the features T1\_L--T5\_L and T1\_R--T5\_R. Notice that according to Table~\ref{tab:perform3} only the (local) $f_0$ component in $\mathcal{F}_0$ is effectively used. From the results it can be concluded
that as the size of the time window decreases, the classification ability of $\mathcal{K}(f_0)$ and $\mathcal{K}(\mathcal{F}_2)$ will equalize.  That is:
\begin{equation}
\label{eq:equal}
\lim_{win.~size\to 0} class.~error^{\Omega}(\mathcal{K}(\mathcal{F}_2)) \approx class.~error^{\Omega}(\mathcal{K}(f_0))
\end{equation}

 Eq.~\ref{eq:equal} characterizes the relationship between watchdog and $\mathcal{F}_2$ based misbehavior detection. It points out that instead of observing {\em each} data packet's delivery in promiscuous mode by the node $s_i$, it can be equally well done in a cooperative way by $s_i$ and $s_{i+2}$, if $win.~size\to 0$.





In other words, if the time window is small enough that it always includes only a single event, the relationship between the events at $s_i$ and $s_{i+2}$ becomes explicit, i.e. a data packet received at $s_{i+2}$ can be unambiguously matched with a data packet forwarded by one of its two-hop neighbors. Decreasing the time window size is however connected with a high communication cost (each packet arrival at $s_{i+2}$ is explicitely reported to $s_i$). For $f_0$ based misbehavior detection, this relationship is straightforward since $s_i$ only evaluates data packets that it just sent. 
 
In terms of learning complexity, the fundamental differences in these two approaches are: 
\begin{itemize}
\item If using $f_0$, the send-overhear relationship is always explicit since $s_i$ can directly observe the forwarding of the data packet that it just sent to $s_{i+1}$. Therefore, the classification task is straightforward.
\item If using $\mathcal{F}_2$ with $win.~size \gg 0$, learning based on data traffic at both $s_i$ and $s_{i+2}$ must be done. Feature averaging over a time window increases the classification task complexity. 
\end{itemize}

The above two observations and Eq.~\ref{eq:equal} offer a rough characterization of the trade-offs between detection approaches executed by a single node and by several nodes in cooperation. 

\begin{table*}[!!!t]
\begin{center}
\begin{tabular}{|c|c|c|c|c|c|c|c|c|}
\hline
\multicolumn {1}{|c|}{} & \multicolumn {2}{|c|}{Dropping} & \multicolumn {2}{|c|}{Delaying} & \multicolumn {2}{|c|}{Wormhole} & \multicolumn {2}{|c|}{Any misbehavior}\\
\hline
Win. size[s] &Det. rate&$CI_{95\%}$&Det. rate&$CI_{95\%}$&Det. rate&$CI_{95\%}$&Det. rate&$CI_{95\%}$\\

\hline
\multicolumn {1}{|c|}{} & \multicolumn {8}{|c|}{$\mathcal{K}(\mathcal{F}_2)$}\\
\hline
50 &94.28 &1.10 &92.89 &0.89 &82.21 &5.91 &93.58 &0.99 \\
100 &91.71 &1.36 &88.53 &1.36 &78.21 &6.85 &90.30 &1.28 \\
250 &88.37 &2.55 &78.65 &1.87 &73.91 &8.48 &84.19 &1.49 \\
500 &84.13 &4.15 &64.64 &3.41 &71.16 &9.01 &76.40 &2.53 \\

\hline
\multicolumn {1}{|c|}{} & \multicolumn {8}{|c|}{$\mathcal{K}(f_0)$}\\
\hline
50 &99.37 &0.17 &99.39 &0.16 &87.59 &5.84 &98.15 &0.62 \\
100 &99.64 &0.22 &99.51 &0.22 &84.15 &6.68 &97.75 &0.71 \\
250 &99.27 &0.40 &99.47 &0.26 &84.49 &4.38 &97.50 &0.66 \\
500 &98.81 &0.67 &99.01 &0.44 &83.63 &5.03 &97.19 &0.85 \\

\hline
\multicolumn {1}{|c|}{} & \multicolumn {8}{|c|}{Co-stimulation: $\mathcal{K}(\mathcal{F}_2) \rightarrow \mathcal{K}(f_0)$}\\
\hline
50 &95.02 &0.92 &93.62 &0.87 &81.00 &6.24 &93.42 &0.72\\
100 &93.53 &1.17 &89.39 &1.51 &75.06 &7.35 &90.00 &0.84\\
250 &89.68 &2.06&79.47 &1.82 &71.09 &8.22 &84.70 &1.06\\
500 &85.82 &3.81 &68.99 &3.47 &69.38 &9.50 &78.89 &1.71\\

\hline
\end{tabular}
\end{center}
\caption{Co-stimulation performance: detection rate.}
\label{tab:perform4}
\end{table*}
\begin{table*}[!!!t]
\begin{center}
\begin{tabular}{|c|c|c|c|c|c|c|c|c|}
\hline
\multicolumn {1}{|c|}{} & \multicolumn {2}{|c|}{Dropping} & \multicolumn {2}{|c|}{Delaying} & \multicolumn {2}{|c|}{Wormhole} & \multicolumn {2}{|c|}{Any misbehavior}\\
\hline

Win. size[s] &~FP rate~~&$CI_{95\%}$&~FP rate~&$CI_{95\%}$&~FP rate~~&$CI_{95\%}$&~FP rate~&$CI_{95\%}$\\

\hline
\multicolumn {1}{|c|}{} & \multicolumn {8}{|c|}{$\mathcal{K}(\mathcal{F}_2)$}\\
\hline
50 &3.44 &0.71 &5.28 &0.76 &8.23 &2.60 &3.55 &0.52 \\
100 &4.40 &0.99 &8.38 &0.91 &6.69 &2.12 &4.65 &0.44 \\
250 &6.22 &1.85 &15.38 &2.06 &16.05 &5.91 &9.28 &1.25 \\
500 &11.08 &3.73 &25.24 &3.56 &19.81 &7.52 &15.09 &2.27 \\

\hline
\multicolumn {1}{|c|}{} & \multicolumn {8}{|c|}{$\mathcal{K}(f_0)$}\\
\hline
50 &0.39 &0.13 &0.78 &0.17 &4.35 &1.67 &0.94 &0.23 \\
100 &0.11 &0.08 &0.61 &0.26 &5.04 &1.81 &0.88 &0.28 \\
250 &0.34 &0.30 &0.49 &0.33 &8.32 &3.04 &1.44 &0.62 \\
500 &0.74 &0.77 &0.62 &0.44 &9.81 &3.75 &1.77 &0.66 \\

\hline
\multicolumn {1}{|c|}{} & \multicolumn {8}{|c|}{Co-stimulation: $\mathcal{K}(\mathcal{F}_2) \rightarrow \mathcal{K}(f_0)$}\\
\hline
50 &0.48 &0.14  &0.45 &0.17 &3.25 &1.51 &0.98 &1.36\\
100 &0.38 &0.19 &0.31 &0.16 &5.14 &2.41 &1.26 &1.75\\
250 &0.97 &0.54 &0.40 &0.21 &3.54 &1.21 &1.28 &1.73\\
500 &1.19 &0.70 &0.44 &0.28 &4.64 &2.13 &1.67 &2.59\\

\hline
\end{tabular}
\end{center}
\caption{Co-stimulation performance: FP rate.}
\label{tab:perform5}
\end{table*}

\subsection{Co-stimulation Based Misbehavior Detection}
\label{sec:costim}

The rule formulated in Eq.~\ref{eq:equal} motivates the following strategy: approximate $\mathcal{K}(f_0)$ with $\mathcal{K}(\mathcal{F}_2)$. If the $\mathcal{K}(\mathcal{F}_2)$ based classification hints a possibility of misbehavior then employ $\mathcal{K}(f_0)$ to get a more reliable prediction. There are two basic reasons for applying this strategy: (i)~computing $\mathcal{F}_2$ is more energy efficient and (ii)~it allows for an energy efficient implementation of the co-stimulation approach presented in Alg.~\ref{fig:detmodel}.

The results achieved by co-stimulation are reported in Tables~\ref{tab:perform4} and \ref{tab:perform5}. The classification performance evaluation of $\mathcal{K}(\mathcal{F}_2)$, $\mathcal{K}(f_0)$ and $\mathcal{K}(\mathcal{F}_2) \rightarrow \mathcal{K}(f_0)$ is based on three separate cross-validation experiments. 
It can be seen that for the four considered time windows sizes, $\mathcal{K}(\mathcal{F}_2)$ and $\mathcal{K}(f_0)$ fulfill the condition formulated in Eq.~\ref{eq:error}. The reported performance, taking into consideration the corresponding $CI_{95\%}$, is also in-line with Eqs.~\ref{eq:det_rate_final} and \ref{eq:fp_rate_final}. Most notably, it can be seen that except for wormholes, the following holds:
\begin{equation}
\label{eq:fp_rate_final2}
FP~rate^{\Omega}_{c_j}(\mathcal{K}(\mathcal{F}_2) \rightarrow \mathcal{K}(f_0)) \cong FP~rate^{\Omega}_{c_j}\mathcal{K}(f_0)
\end{equation}

The results show that for $win.~size = 50s$, $18.2\%$ and $13.3\%$ wormholes were misclassified as belonging to the ``normal'' class, when $\mathcal{K}(\mathcal{F}_2)$ and $\mathcal{K}(f_0)$ were applied, respectively. These two factors contributed to this result: (i) a wormhole start node can appear to ``lose'' a variable number of data packets depending on the wormhole utilization in a given time window , and (ii) the used features were not sufficient for this type of misbehavior. 


\subsection{Analysis of Co-stimulation Energy Efficiency}
\label{sec:costim-energy}

The rationale for applying co-stimulation rests on its ability to stimulate energy efficiency. Let us now therefore formulate an energy model for co-stimulation.

Unlike when $\mathcal{K}(f_0)$ gets exclusively applied, in our approach $\mathcal{K}(f_0)$ will get used only if (i) a true positive was detected by $\mathcal{K}(\mathcal{F}_2)$ or (ii) a false positive was mistakenly detected by $\mathcal{K}(\mathcal{F}_2)$. This means, for a {\em misbehavior free} ad hoc network, the energy saving over an exclusive $\mathcal{K}(f_0)$ approach is related to $FP~rate^{\Omega}_{c_j}(\mathcal{K}(\mathcal{F}_2))$. We focus on the energy efficiency analysis in a misbehavior free ad hoc network, since it is reasonable to assume that an ad hoc network will work reliably, most of the time.  

With respect to the above said, we assume the following energy model for co-stimulation in a misbehavior free ad hoc network:
\begin{equation}
\label{eq:costimulation}
\xi(n) = \xi_{\mathcal{F}_2} + FP~rate^{\Omega}_{c_j}(\mathcal{K}(\mathcal{F}_2)) \times \xi_{f_0}(n)
\end{equation} 

where $n$ is the number of data packets that need to be overheard by $s_i$ in promiscuous mode for the purpose of $f_0$ based classification.  $\xi(n)$ is the total energy consumption after co-stimulation in a time window of the length $win.~size$, $\xi_{\mathcal{F}_2}$ is the energy consumption related to the computation of $\mathcal{F}_2$ and $\xi_{f_0}(n)$ is the energy consumption related to the computation of $f_0$. In the following, we also assume that $c_j =$ ``any misbehavior''. Additionally, we assume that decision tree query costs are negligible, if compared to feature computation costs that are dominated by the communication costs. 

The choice of $n$ depends on the expected data traffic pattern as well as the type of misbehavior. With respect to our experimental setup, we assume that $\mathcal{K}(f_0)$ is based on a 50-second time window size.  Note that if $s_{i+1}$ forwards data packets for a single connection then $n = 0.5 \times win.~size$ under our data traffic model, i.e. we consider $n = 25$.

\begin{table*}[!!!tp]
\begin{center}

\begin{tabular}{|c|c|c|c|c||c|c|}
\hline
$win.~size$ & $FP~rate$ [\%]& $\xi(25)$ & $\xi'(25)$ &$\gamma(25)$ & Det.rate [\%] & FP rate [\%]\\
  $[s]$ & $\mathcal{K}(\mathcal{F}_2)$ & [mJ] & [mJ] & [\%]&  $\mathcal{K}(\mathcal{F}_2) \rightarrow \mathcal{K}(f_0)$ &$\mathcal{K}(\mathcal{F}_2) \rightarrow \mathcal{K}(f_0)$\\
\hline
50   &3.55 &2.33 & 23.29 & 82.73 &93.42 &0.98 \\
100  &4.65  &2.48 &12.39 & 90.81 &90.00 &1.26 \\
250  &9.28  &3.10 &6.20 & 95.40 &84.70 &1.28 \\
500  &15.09  &3.89 & 3.89 & 97.12  &78.89 &1.67\\

\hline

\end{tabular}
\end{center}
\caption{Co-stimulation: energy consumption and misbehavior detection performance.}
\label{tab:energy}
\end{table*}

{\em Energy Consumption Analysis}: Feeney and Nilsson investigated in~\cite{feeney2001iec} the energy consumption of a Lucent 2Mbps IEEE802.11 wireless card. They concluded that energy consumption can be modeled as $a \times size + b$, where $size$ is the data packet size in bytes. The constants $a$ and $b$ reflect the consumption  in $\mu J$ when sending ($a=1.9, b=454$), receiving ($a=0.5, b=356$) or overhearing ($a=0.39,b=140$) a data packet. This translates into the following energy models for $\mathcal{K}(f_0)$ and $\mathcal{K}(\mathcal{F}_2)$, respectively:
\begin{equation}
\label{eq:f0}
\xi_{f_0}(n) = n \times (0.39 \times size(data) + 140)
\end{equation} 
\begin{equation}
\label{eq:F2}
\xi_{\mathcal{F}_2} = 2 \times ((1.9 \times size(\hat{f}_2) + 454) + (0.5 \times size(\hat{f}_2) + 356))
\end{equation} 
where $size(data)$ is the data packet size in bytes and $size(\hat{f}_2)$ is the size of the data packet carrying $\hat{f}_2^{s_{i+2}}$ in bytes. Eq.~\ref{eq:f0} reflects the fact that in promiscuous mode the energy consumption grows linearly with the number of data packets overheard. Eq.~\ref{eq:F2} reflects the cost of sending $\hat{f}_2^{s_{i+2}}$ over two hops to $s_{i}$.  



The results for energy efficiency using co-stimulation are shown in Table~\ref{tab:energy}. We applied the following parameters: $size(data) = 1kB$ and $size(\hat{f}_2) = 48B + size(header)$ (corresponding to 24 features, 2 bytes reserved for each feature; $size(header) = 0$, if $\hat{f}_2^{s_{i+2}}$ is transported using piggybacking). $\xi'(n)$ and $\gamma(n)$ are defined as follows:

\begin{equation}
\label{eq:gamma}
\xi'(n) = \frac{500.0}{win.~size} \times \xi(n)~~~~~~~\gamma(n) = 1.0 - \frac{\xi'(n)}{\xi_{f_0}(250)}
\end{equation} 

$\xi'(n)$ is the adjusted energy consumption for a 500-second time period. $\gamma(n)$ is the energy saving relative to $\xi_{f_0}(250)$, i.e. relative to the energy consumption of $f_0$ based classification in a 500-second time window:

Inspecting Table~\ref{tab:energy}, it can be seen that under our experimental setup co-stimulation can save up to $97.12\%$ energy. It can also be seen that co-stimulation offers a possibility to choose a trade-off between energy consumption and detection performance. This can be seen by comparing $\gamma(25)$ and the right-hand side of Table~\ref{tab:energy}, where the classification results for the ``any misbehavior'' class are shown. 

\begin{figure*}[!!!tp]
\begin{center}
  {
    \subfigure[Energy consumption vs time.]{
    \epsfig{file= 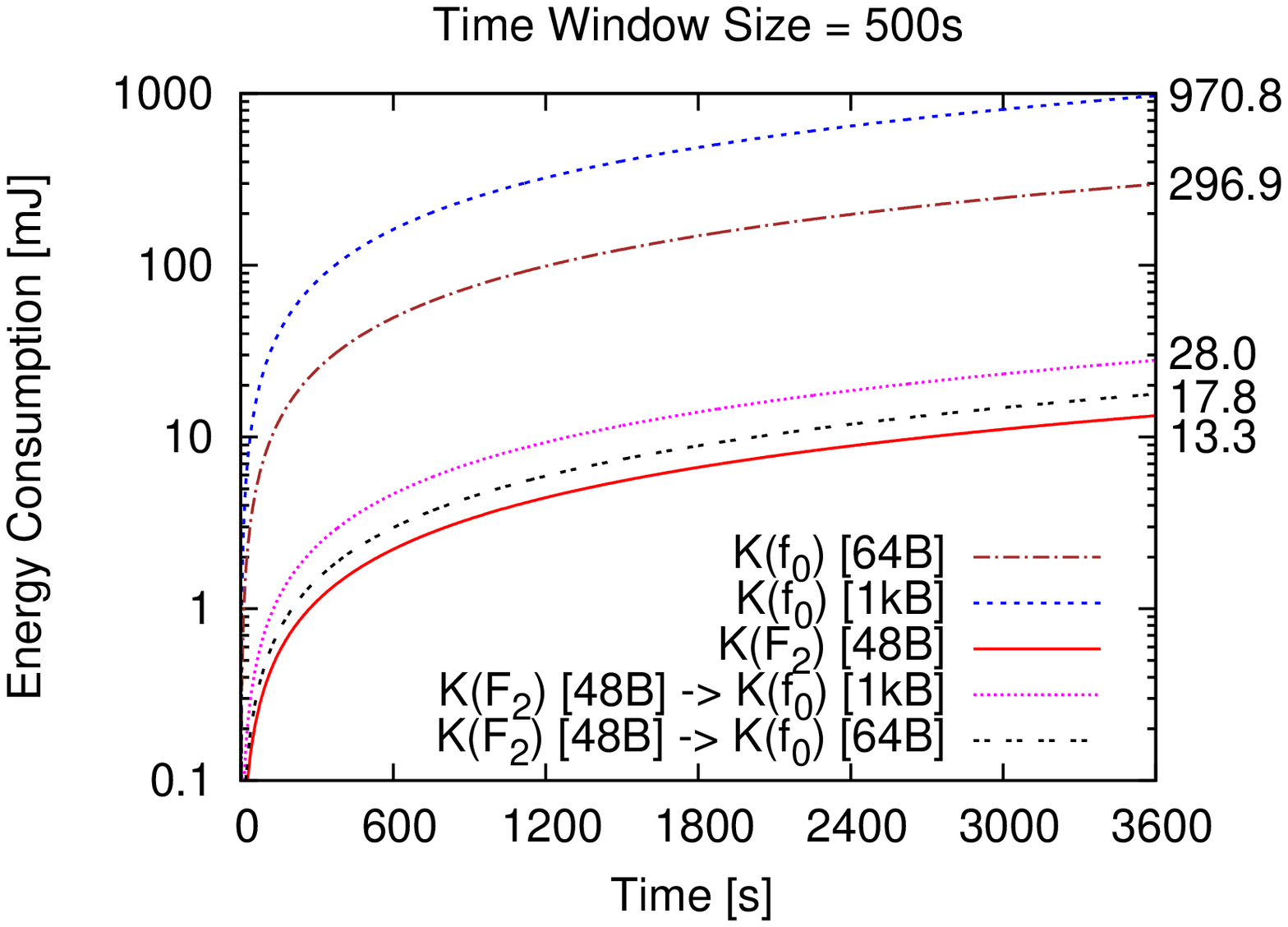, width=0.45\linewidth}}
    \subfigure[Energy consumption vs $FP~rate(\mathcal{F}_2)$.]{
    \epsfig{file= 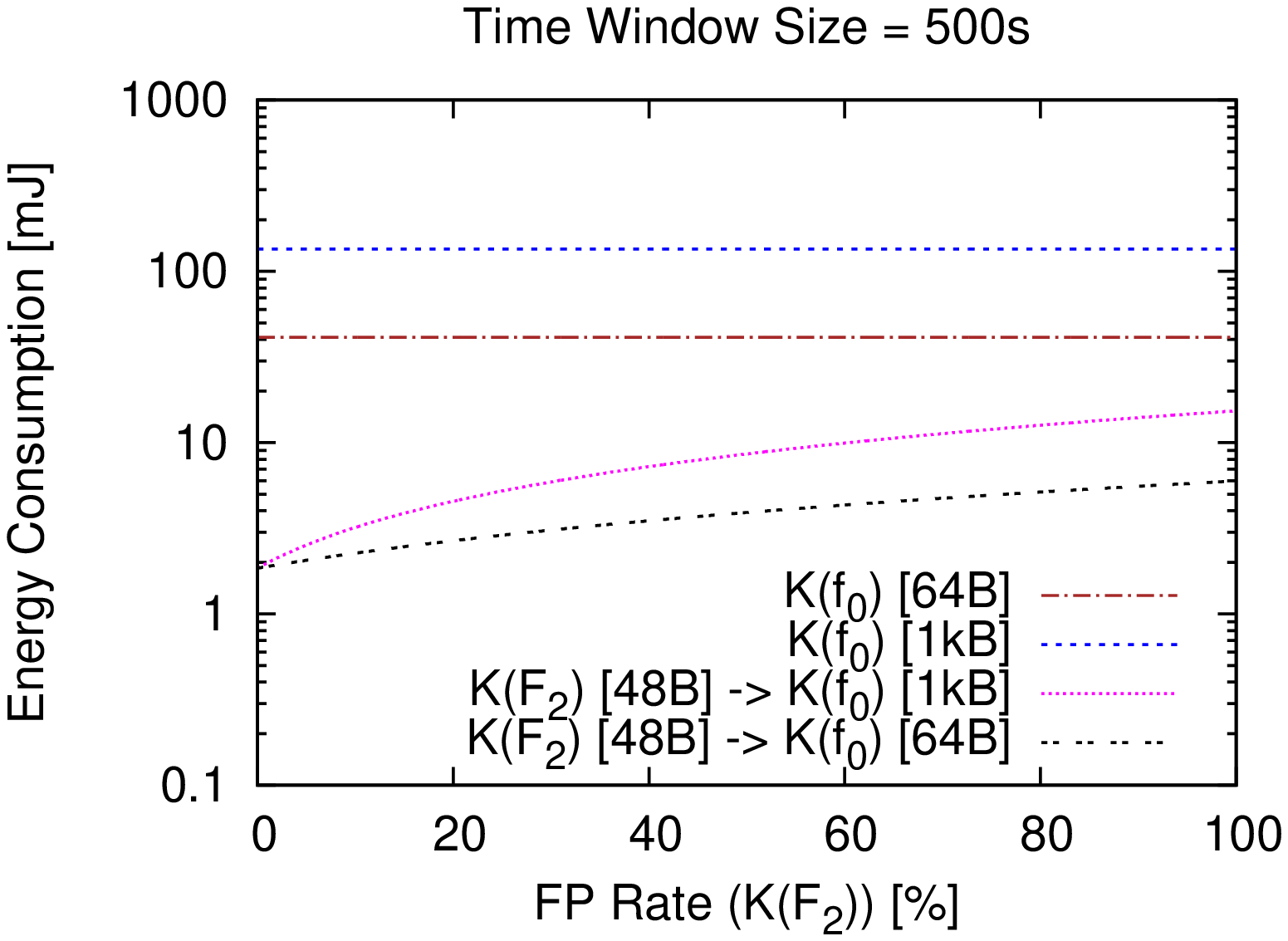, width=0.45\linewidth}}
  }
\caption{Energy consumption.}
\label{fig:energy}
\end{center}
\end{figure*}

In Fig~\ref{fig:energy}(a), the accumulated energy consumption using 500-second time windows is shown. It can be seen that after 60 minutes of operation, the energy saving in a misbehavior free ad hoc wireless network for the data packet size of 1kB is nearly two orders of magnitude. For a smaller data packet size of $64$ bytes, more typical for wireless sensor networks, the energy saving is more than one order of magnitude. In Fig~\ref{fig:energy}(b) the dependence of energy consumption on $FP~rate~(\mathcal{F}_2)$ is shown.

By solving $\xi_{f_0}(n) = \xi_{\mathcal{F}_2}$,  we can show that $\xi_{\mathcal{F}_2} < \xi_{f_0}(n)$, if $n > 3.43$, i.e. the condition formulated in Eq.~\ref{eq:energy} is satisfied for a reasonable time window size. 


\subsection{Further Discussion of Results}

We restricted ourselves to three basic types of misbehavior. Our goal was to investigate one representative from the qualitative group of misbehavior types (packet dropping), one from the quantitate group (packet delaying) and one from the topology group (wormholes). Many types of misbehavior can be classified within these three groups. For example, a data packet manipulation attempt falls within the qualitative group since it can be detected by monitoring the data packet size and/or correction code, learning the usual manipulation rate (due to e.g. a routing protocol) and subsequent classification.

Our approach to misbehavior detection can be considered very general. It is well known that some types of misbehavior can be efficiently detected, if for example time synchronization or GPS (Global Positioning System) information is available; see e.g. results on packet leashes for wormhole detection due to Hu et al.~\cite{hu2006waw}. Such specialized capabilities would however interfere with our intention to contrast implications of a non-cooperative versus cooperative detection ($\mathcal{K}(f_0)$ vs co-stimulation based misbehavior detection).

The results presented in Tables~\ref{tab:perform}, \ref{tab:perform2} and \ref{tab:perform4} offer a good performance guidance for detection systems that aim at profiting from traffic measurements in the neighborhood of a misbehaving node. Even though misbehavior detection can be also done by an independent third-party node that is not lying on the given connection but that can easily overhear the transmission, its observations with respect to the misbehaving node are not going to be more precise than the combined information from nodes $s_i$ and $s_{i+2}$. Such a scenario also implicates a continuous operation in promiscuous mode. In addition to energy inefficiency, operation in promiscuous mode increases vulnerability to intrusions since once a data packet gets intercepted and (temporarily) stored in the memory, it allows for an application of the techniques described in~\cite{szor2005acv}. This can lead to execution of the code carried in the packet's data portion. Unlike in usual packet forwarding, (i) such a scenario would simplify choosing a specific victim and (ii) it would be harder to detect because other nodes might be unaware of the fact that this node was able to overhear the transmission.

We concentrated on features that can be computed without much computational overhead. Our results could have been somewhat different if a more complex Fourier or wavelets analysis of the packet stream had been done. As the results by Barford et al. point out~\cite{barford2002san}, this could lead to good anomaly detection rates. 

The classification in our setup is done on a single time window basis. This has a certain energy efficiency effect as a post-processing phase (statistical analysis, clustering) can be avoided. This positively impacts the time to detect a misbehavior.

The BIS is an inherently distributed system with functionality that can be very hard to mimic. For example, the success of the negative selection, a learning mechanism applied in training and priming of T-cells in the thymus, rests on the efficiency of the blood-thymic barrier that guarantees that the thymus stays pathogen free at all times. This implies that T-cells being trained in the thymus never encounter any type of pathogen before released into the body. This helps tremendously in detecting foreign cells. Mapping functionality of the BIS to a computational paradigm is a hot topic within the AIS community. Our goal was to concentrate on the simplest mechanisms. Most notably, we were motivated by the interplay between the innate and adaptive immune system.

The mechanisms of the innate immune system bear a certain resemblance to the $\mathcal{F}_0$ based classification approach. The innate system is for example very efficient in signaling tissue injury or damage to the adaptive immune system. This ability, as pointed out before, has been induced over the evolutionary time but is based on some very rudimentary methods such as recognizing an unusually high level of dead or damaged self cells (e.g. blood cells). This can be directly compared with the very straightforward functionality of watchdogs. Similarly, the more machine learning extensive classification approach based on $\mathcal{F}_2$ can be, in our opinion, compared with the adaptive immune system. 

\section{Conclusions}
\label{sec:conclusions}
We presented and experimentally evaluated a novel immuno-inspired energy efficient approach to misbehavior detection in ad hoc wireless networks. We demonstrated that under our experimental setup, the (communication related) energy saving  for 1kB data packets, if compared to watchdog monitoring, is nearly two orders of magnitude. For smaller data packets of 64 bytes, more typical for wireless sensor networks, the energy saving is above one order of magnitude. 

We achieved a good control of the false positives rate. This is important in order to reduce the maintenance costs of ad hoc networks.  We also showed that our co-stimulation approach becomes cheaper than its watchdog based counterpart at a time window size of 6.86 second. 

We demonstrated a relationship between the energy efficiency and detection performance. This approach can be thus used to find a trade-off between misbehavior detection performance and energy efficiency.

We applied three different node misbehavior models: data packet dropping, data packet delaying and wormholes. A wormhole can only be formed if two nodes closely cooperate, i.e. it is an instance of misbehavior done in collusion. 

We would like to note that our results are independent from the global network topology, since in our approach we only looked at a two-hop segment of a connection. Since we evaluated the performance at 20 distinct nodes with different node degrees, our results provide a robust estimate of the expected performance.

Motivated by the results presented herein, Drozda et al. investigated an error propagation algorithm~\cite{drozda2009innate} that takes advantage of Eq.~\ref{eq:equal} in order to induce systemic resistance against misbehavior. It also removes the reliance on a labeled dataset for learning the normal behavior and misbehavior.

The datasets used in our experiments are available for download~\cite{datasets}.

\section*{Acknowledgments}

This work was supported by the German Research Foundation (DFG) under the grant no. SZ 51/24-2 (Survivable Ad Hoc Networks -- SANE).

\balance


\end{document}